# Characterizing the speed and paths of shared bicycles in Lyon

Pablo Jensen (a,c), Jean-Baptiste Rouquier (b), Nicolas Ovtracht (c) and Céline Robardet (d)

(a) Institut des Systèmes Complexes Rhône-Alpes (IXXI) ; Laboratoire de Physique, École Normale Supérieure de Lyon, 69007 Lyon, France
(b) ISC PIF (CNRS and CRÉA, École Polytechnique, Paris, France
(c) Laboratoire d'Économie des Transports, Université Lyon-2, Lyon, France;
(d) Université de Lyon, INSA-Lyon, CNRS, LIRIS UMR5205, F-69621 France

**Abstract :**
Thanks to numerical data gathered by Lyon's shared bicycling system *Vélo'v*, we are able to analyze 11.6 millions bicycle trips, leading to the first robust characterization of urban bikers' behaviors. We show that bicycles outstrip cars in downtown Lyon, by combining high speed and short paths.These data also allows us to calculate *Vélo'v* fluxes on all streets, pointing to interesting locations for bike paths.

Lyon's *Vélo'v* started in May 2005 and represented the first massive shared bicycling system. Today, 4000 bikes can be taken in one of the 343 stations spread across the city (Figure 1). On average, some 16 000 journeys per day are completed, but when public transportation is on strike, this number doubles. Our dataset, kindly provided by *Vélo'v* operator JC Decaux [Decaux, 2010], contains all the trips that occurred between May 25$^{th}$ 2005 and December 12$^{th}$ 2007. Each record contains the location and time of the beginning and the end of the trip, as well as the precise trip distance provided by a counter on the bike. The average trip distance is 2.49 km and the average trip time 14.7 min.

Speed is an essential feature to determine the efficiency of transportation systems. Figure 2a shows that bikers' *average* speed reaches a peak of 14.5 km/h in the early weekday mornings, when there is almost no hindrance by cars or traffic lights. Interestingly, this early morning peak is the only moment when average summer and winter speeds (not shown) are identical, pointing to an intrinsic limit of the system. For any other moment of the day, winter speeds are higher (up to 9\% in the evening). However, experienced or hurried bikers can go much faster : the top 10\% (black curve in Figure 2a) reach 20 km/h. Lower speeds can be explained by traffic conditions or lack of hurry. The former explain the morning speed decrease, even if bikers may speed up if needed, as shown by the increase of average speed at rush hours (especially 8:45) all five working days (Figure 2a, inset) and the clear peaks in the fastest bikers' speed (7:45 and the plateau 8:45 to 9:45). When there is no need to hurry, average speed falls to 10 km/h, as in weekends afternoons. Intriguingly, wednesday morning speeds (upper curve in the inset) are systematically higher than other weekdays. Since car traffic does not decrease on wednesdays, we speculate that this higher speed might be related to a higher proportion of (faster) masculine bikers, since a significant fraction of women stay home to care for children on wednesdays [Bel, 2008].

The data allows us to compare the *Vélo'v* real trip distances between stations to car and pedestrian distances. Figure 2b shows that bike's paths are much closer to pedestrians'. Actually, when there exists a shortcut, 68.2\% of bike's trips are shorter than car's (left of the vertical line in Figure 2b), the average distance reduction being 13\%. The proportion is slightly weaker (61.3\% out of 3,506,294 trips) if all weekday hours are taken into account. In both cases, to avoid artifacts, we only keep trips longer than 500m between a couple of stations where car distance is significantly longer than pedestrians' (at least 200m). This suggests that, as specific bicycle tracks were virtually unknown in Lyon at that time, most bikers use sidewalks, drive the wrong way up one-way streets or use the bus / tramway lanes. Finally, using data on the fluxes between couples of stations and assuming that bikers follow pedestrian paths, we can calculate *Vélo'v* fluxes on all streets (Figure 1). Knowing these fluxes should help town authorities in creating bike paths where they are most needed.

Our analysis shows that in morning rush hours, bike's average speed - in real conditions and for average users - is 13.5 km/h. Adding the effect of shorter bike's trips (Figure 2b) leads to an effective speed close to 15 km/h. This speed should be compared to car's average speeds in downtown European cities, which vary from 10 km/h to 15 km/h [Paris, 2010; Prudhomme and Bocarejo, 2005]. Therefore, even when considering bare speeds, bikes are faster than cars in downtowns. This advantage is not cancelled by including in the speed calculations the average time needed to reach the car or the bike. Indeed, even if walking distance to car's parking is difficult to estimate, it is of order 200m in downtown Lyon, which is similar to the distance to the closest *Vélo'v* station. On the other hand, finding a parking slot is generally more difficult than an empty *Vélo'v* slot in downtown areas. The combination of these elements may explain why, after *Vélo'v* gave a modern look to cycling in 2005, the number of bikers has almost doubled in Lyon [Grand Lyon, 2010].

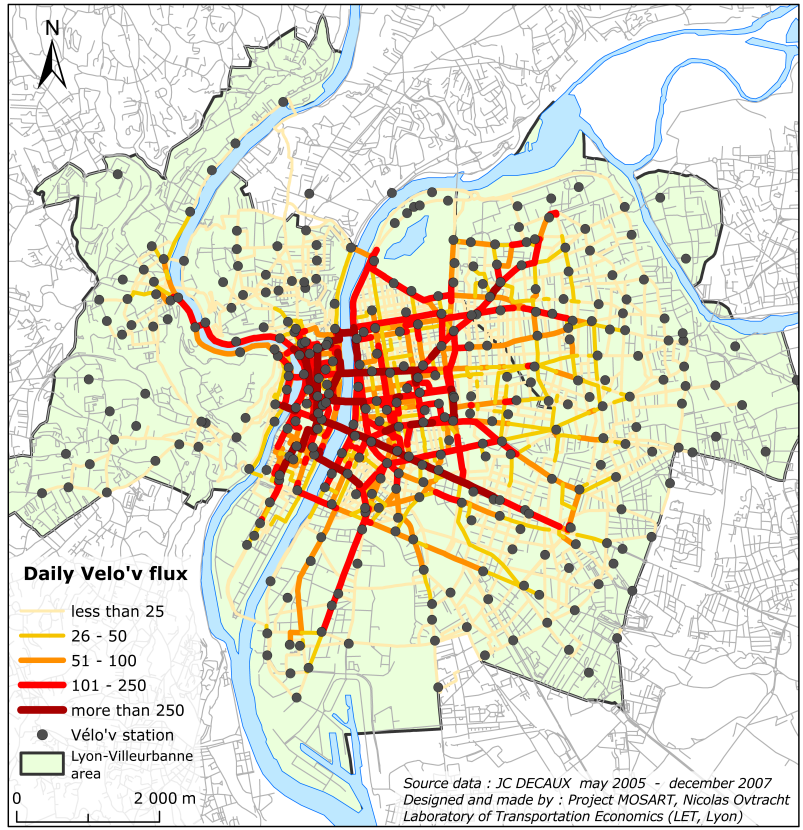

FIG. 1: Map of *Vélo'v* stations and daily average fluxes on Lyon's streets. The map was obtained thanks to the *MOSART* modelling platform (http://mosart.let.fr/)

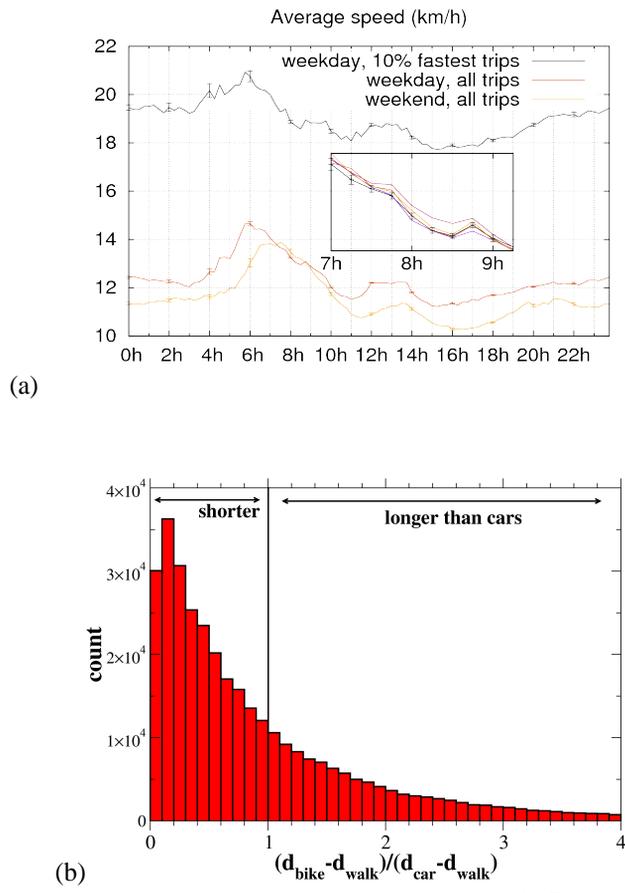

FIG. 2: **(a)** Average speed vs hour in weekdays and weekends. Bars give the 95\% confidence interval. **(b)** Histogram of (velov - pedestrian) / (car - pedestrian) distances for the 375,165 morning (7 to 9am) weekday trips, when bikers are likely to have time constraints.